\RequirePackage{lineno}
\documentclass[aps,prc,reprint,showpacs]{revtex4-1}
\usepackage{amsmath}
\usepackage{graphicx}
\usepackage{dcolumn}

\begin{document}
%\linenumbers

\title{Calculating Jet $v_n$ and the Event Plane in the Presence of a Jet}
\author{Alice Ohlson}
\affiliation{Yale University, New Haven, Connecticut 06511, USA}
\date{\today}

\begin{abstract}
Advances in measurements of jets and collective phenomena in ultrarelativistic heavy ion collisions have led to further understanding of the properties of the medium created in such collisions.  Measurements of the correlations between the axes of reconstructed jets and the reaction plane or second-order participant plane of the bulk medium (defined as jet $v_2$), as well as the higher-order participant planes (jet $v_n$), provide information on medium-induced parton energy loss.  Additionally, knowledge of jet $v_n$ as well as the ability to reconstruct the event plane in the presence of a jet are necessary in analyses of jet-triggered particle correlations, which are used to study medium-induced jet shape modification.  However, the presence of a jet can bias the event plane calculation, leading to an overestimation of jet $v_2$.  This paper proposes a method for calculating jet $v_2$ (and by extension, the higher jet $v_n$ harmonics) and the event plane in an unbiased way, using knowledge of the azimuthal angle of the jet axis from full jet reconstruction.  
\end{abstract}
\maketitle

\section{Introduction \& Motivation}
The strongly coupled medium produced in high-energy collisions of large nuclei at the Relativistic Heavy Ion Collider (RHIC)~\cite{STARwhite, PHENIXwhite, PHOBOSwhite, BRAHMSwhite} and at the Large Hadron Collider (LHC)~\cite{ALICEqm, CMSqm, ATLASqm} exceeds the energy density at which quarks and gluons are expected to be deconfined, and is known as the quark-gluon plasma (QGP)~\cite{lqcdRischke}.  The properties of the QGP are studied through many diverse approaches, including analyses of (a) collective particle behavior and (b) medium-induced parton energy loss~\cite{RHICreviewJacobs}.  

\subsection{Collective Flow \& Geometry}
Most particles produced in heavy ion collisions result from the hadronization of the QGP, and are collectively correlated with respect to the initial geometry of the collision in a way that suggests that the QGP exhibits hydrodynamic flow on the parton level~\cite{partonicFlow}.  

%The geometry of the colliding nuclei can be described by the ``reaction plane'', the plane defined by the impact parameter and the beam direction.  However, the region defined by the distribution of the participating nucleons (the ``participant zone'') is not perfectly described by the geometrical overlap of the colliding nuclei.  The $n^{\text{th}}$-order planes of symmetry of the participant zone are known as the $n^{\text{th}}$-order ``participant planes.''  
The geometry of the colliding nuclei can be described by the ``reaction plane,'' the plane defined by the impact parameter and the beam direction.  However, the positions of the participating nucleons fluctuate event-to-event so that the matter distribution and its symmetries cannot be fully characterized by the geometrical overlap of the colliding nuclei. For this reason, the concept of participant planes defined by the $n^{\text{th}}$-order symmetries of the participating nucleons was introduced~\cite{firstParticipantPlane,Broniowski}.

As the QGP expands, pressure gradients boost the particles along the participant planes, and the coordinate-space eccentricity is converted into an anisotropy in momentum-space~\cite{flowReview}.  The azimuthal angular ($\phi$) distribution of the particles can be expanded in Fourier coefficients with respect to the azimuthal angle of the reaction plane ($\Psi_{\text{RP}}$), as shown in Eqn.~(\ref{eq:fourier}), or any order participant plane ($\Psi_{\text{PP,}m}$), as in Eqn.~(\ref{eq:fourierpp})~\cite{fourier}:   
\begin{linenomath}\begin{equation}\label{eq:fourier}
\frac{\mathrm{d}N}{\mathrm{d}\left(\phi-\Psi_{\text{RP}}\right)} \propto 1+\sum^{\infty}_{n=1}2 v_n \cos\left[n\left(\phi-\Psi_{\text{RP}}\right)\right]
\end{equation}\end{linenomath}
\begin{linenomath}\begin{equation}\label{eq:fourierpp}
\frac{\mathrm{d}N}{\mathrm{d}\left(\phi-\Psi_{\text{PP,}m}\right)} \propto 1+\sum^{\infty}_{n=1}2 v_n \cos\left[n\left(\phi-\Psi_{\text{PP,}m}\right)\right]
\end{equation}\end{linenomath}

In semicentral events the participant zone is roughly elliptical in coordinate space, because the geometrical overlap region of the colliding nuclei is almond-shaped; the minor axis of the ellipse is the $2^{\text{nd}}$-order participant plane, which is closely aligned with the reaction plane.  Due to the elliptical shape of the interaction region, the $v_2$ term is dominant, in all but the most central collisions, when measured with respect to either $\Psi_{\text{RP}}$ or $\Psi_{\text{PP},2}$.  Recent theoretical~\cite{firstFlow, Paulv3, alver} and experimental~\cite{PHENIXv3, ALICEv3} work has shown that fluctuations in the initial state can also result in nonzero higher-order $v_n$ coefficients, in particular $v_3$, when calculated with respect to the $n^{\text{th}}$-order participant plane.  Mixed harmonics, such as $v_3$ with respect to $\Psi_{\text{PP,}2}$, will not be discussed in this paper.  

Both the reaction plane and the participant planes are theoretical concepts, while the experimental observable is the $n^{\text{th}}$-order event plane (denoted by $\Psi_{\text{EP},n}$).  Depending on the method used to the reconstruct the $2^{\text{nd}}$-order event plane, the result may be more closely related to the $2^{\text{nd}}$-order participant plane or to the reaction plane.  For example, it is suggested that asymmetries in the distribution of spectator neutrons detected at forward pseudorapidities are more sensitive to the reaction plane geometry, while the asymmetries of produced particles at midrapidity are more sensitive to the participant geometry~\cite{Bhalerao}.  

\subsection{Jets \& Jet Quenching}
Partons resulting from hard scatterings in the initial stages of heavy ion collisions traverse the QGP and fragment into tightly correlated ``jets'' of particles.  The production and fragmentation of these partons is well described by perturbative QCD in proton-proton ($pp$) collisions~\cite{pQCD}, where no QGP is expected to be formed.  Therefore by measuring the modification of these jets in heavy ion collisions one can learn about parton interactions with the strongly coupled medium.  Suppression of particles with high transverse momentum ($p_{\text{T}}$) has been observed experimentally in measurements of $R_{\text{AA}}$~\cite{STARraa, PHENIXraa, ALICEraa, CMSraa} and in two-particle correlations~\cite{STARdihadron, STARdihadron2, STARdihadron3, PHENIXdihadron, PHENIXdihadron2, ALICEdihadron, CMSdihadron}, and is often attributed to medium-induced parton energy loss.  

A few analyses have been performed which merge the studies of jets and flow, such as high-$p_{\text{T}}$ $v_2$~\cite{ALICEhighptv2, CMShighptv2} and dihadron correlations with respect to the event plane~\cite{STARhighptRP, PHENIXdihadronRP}.  While these methods use high-$p_{\text{T}}$ particles as proxies for jets, recent advances in jet reconstruction techniques~\cite{fastjet2} have made it possible to study highly energetic jets in heavy ion collisions directly.  

\subsection{Jet $v_n$}
It is expected that partons produced in hard scatterings are emitted isotropically in the plane transverse to the beam direction, independent of the initial QGP geometry.  However, it is likely that medium-induced parton energy loss depends on the length of the parton's path through the QGP, and should depend on the QGP geometry.  Pathlength-dependent jet suppression can give rise to a difference in the number of jets reconstructed in the direction of the event plane and out of the event plane, depending on the jet $p_{\text{T}}$ range and the parameters used in jet reconstruction, such as $p_{\text{T}}$ cuts and the resolution parameter $R$.  This effect would result in a correlation between reconstructed jets and the $n^{\text{th}}$-order participant planes, leading to $v_n$ coefficients in a Fourier series representation of the azimuthal distribution of the reconstructed jet axes ($\Psi_{\text{jet}}$):

\begin{linenomath}\begin{equation}\label{eq:fourierJet}
\frac{\mathrm{d}N}{\mathrm{d}\left(\Psi_{\text{jet}}-\Psi_{\text{PP,}m}\right)} \propto 1+\sum^{\infty}_{n=1}2 v_n^{\text{jet}} \cos\left[n\left(\Psi_{\text{jet}}-\Psi_{\text{PP,}m}\right)\right]
\end{equation}\end{linenomath}

Jet $v_n$ (denoted by $v_n^{\text{jet}}$) describes the correlation between the jet axis and the reaction plane or participant planes, and should not be confused with a measure of the correlation of jet fragments.  Non-zero jet $v_n$ does not necessarily indicate that jets experience hydrodynamic flow, since pathlength-dependent energy loss is a more likely physical explanation.  ``Jet $v_n$,'' which describes a correlation between the jet and the collision geometry, should be distinguished from pressure-driven ``jet flow,'' which imposes a physical interpretation on any such correlation.  

Pathlength-dependent jet modification could also manifest itself as a modification of the jet shape (associated particle yield and/or width of the jet peaks) as a function of the relative angle between the jet axis and the event plane.  This effect could potentially be seen in an analysis of jet-triggered particle correlations with respect to the event plane.  In order to perform such an analysis, it is necessary to reconstruct the event plane accurately in the presence of a jet.  Additionally, for such correlation studies, knowledge of jet $v_n$ is crucial for background subtraction purposes.  

\section{Jet -- Event Plane Bias}
Unfortunately, calculating jet $v_n$ is not as simple as reconstructing the event plane using the standard technique (Eqn.~(\ref{eq:EP})) and then calculating $v_n^{\text{jet}}\{\text{EP}\}$ using Eqn.~(\ref{eq:v2std}):
\begin{linenomath}\begin{equation}\label{eq:EP}
\Psi_{\text{EP,}n} = \frac{1}{n} \tan^{-1} \left(\frac{\sum_i w_i\sin\left(n\phi_i\right)}{\sum_i w_i\cos\left(n\phi_i\right)}\right)
\end{equation}
\begin{equation}\label{eq:v2std}
v_n^{\text{jet}}\{\text{EP}\} = \frac{\langle \cos\left(n(\Psi_{\text{jet}}-\Psi_{\text{EP,}n})\right)\rangle}{\langle \cos\left(n(\Psi_{\text{EP,}n}-\Psi_{\text{PP,}n})\right)\rangle}
\end{equation}\end{linenomath}
(In Eqn.~(\ref{eq:EP}) the index $i$ runs over all particles and the weights $w_i$ are chosen to maximize the event plane resolution.)

If jet fragments are included in the event plane calculation it will cause the reconstructed event plane to be biased towards the jet axis.  While the bias may be small on an event-by-event basis, it can cause a significant overestimation of the jet $v_n$.  Therefore, a method for removing jet particles from the event plane calculation is desirable.  Several ideas for removing the jet particles from the event plane reconstruction exist, such as removing all the tracks within a cone around the jet axis from the calculation, or introducing a pseudorapidity ($\eta$) gap between the jet and the particles used to reconstruct the event plane.  The former idea cannot remove all particles associated with a jet for reasonable cone sizes, and introduces further correlations that contribute to the artificial jet -- event plane bias.  The latter idea requires reconstructing the event plane at large pseudorapidities which not only decreases the event plane resolution, but also requires a detector with good angular resolution at forward angles.  
%While these detector capabilities are present in ATLAS and CMS, for detectors with limited acceptances such as ALICE and STAR it is desirable to calculate the event plane at mid-rapidity.  

In this paper a method is presented for calculating jet $v_2$ and the $2^{\text{nd}}$-order event plane by accounting for the presence of a high-$p_{\text{T}}$ jet, instead of attempting to remove the jet particles from the event plane reconstruction.  The method is then extended to calculate $v_n^{\text{jet}}$ and $\Psi_{\text{EP,}n}$.  

\section{The Standard Event Plane Calculation}
The standard method of calculating the event plane~\cite{standardEP} is by defining a flow vector, $\mathbf{Q}_2$, that describes the bulk particle distribution.  The components of $\mathbf{Q}_2$ are shown in~(\ref{eq:Q0}), where the index $i$ again runs over all the particles used in the event plane reconstruction.  The vector $\mathbf{Q}_2$ and the beam axis define the event plane.  
\begin{linenomath}\begin{align}\label{eq:Q0}
Q_{2,x} = \sum_i w_i\cos(2\phi_i) = Q_2\cos(2\Psi_{\text{EP,}2})\\
Q_{2,y} = \sum_i w_i\sin(2\phi_i) = Q_2\sin(2\Psi_{\text{EP,}2})\nonumber
\end{align}\end{linenomath}
Events containing jets, however, can be decomposed into two vectors: one describing the bulk distribution ($\mathbf{Q}_2$) and one describing the jet constituents ($\mathbf{A}_2$).  When a jet is present, the standard event plane method does not find the angle of $\mathbf{Q}_2$ but rather the angle of the sum of these two vectors ($\mathbf{G}_2 = \mathbf{Q}_2+\mathbf{A}_2$), as shown in~(\ref{eq:G0}):  
\begin{linenomath}\begin{align}\label{eq:G0}
G_{2,x,\text{lab}} &= \sum_i w_i\cos(2\phi_i) \\
&= \sum_{i\in\text{bulk}} w_i\cos(2\phi_i) + \sum_{i\in\text{jet}} w_i\cos(2\phi_i) \nonumber\\
&= Q_2\cos(2\Psi_{\text{EP,}2}) + A_2\cos(2\Psi_{\text{jet}})\nonumber\\
G_{2,y,\text{lab}} &= \sum_i w_i\sin(2\phi_i) \nonumber\\
&= \sum_{i\in\text{bulk}} w_i\sin(2\phi_i) + \sum_{i\in\text{jet}} w_i\sin(2\phi_i) \nonumber\\
&= Q_2\sin(2\Psi_{\text{EP,}2}) + A_2\sin(2\Psi_{\text{jet}})\nonumber
\end{align}\end{linenomath}
The difficulty is to untangle $\mathbf{A}_2$ and $\mathbf{Q}_2$ in order to extract the angle of $\mathbf{Q}_2$ alone.   Assuming that jets can be accurately reconstructed in heavy ion collisions, there is another piece of information that is known: the angle of $\mathbf{A}_2$, which is the jet axis.  

\section{A New Event Plane Method}
If $G_{2,x}$ and $G_{2,y}$ are calculated with respect to the jet axis (instead of in the laboratory frame, as shown in~(\ref{eq:G0})), then $A_2$ only appears in one of the terms:  
\begin{linenomath}\begin{align}
G_{2,x,\text{jet}} &= \sum_i w_i\cos(2(\phi_i-\Psi_{\text{jet}})) \\
&= Q_2\cos(2(\Psi_{\text{EP,}2}-\Psi_{\text{jet}})) + A_2\nonumber\\
G_{2,y,\text{jet}} &= \sum_i w_i\sin(2(\phi_i-\Psi_{\text{jet}})) \nonumber\\
&= Q_2\sin(2(\Psi_{\text{EP,}2}-\Psi_{\text{jet}})) \nonumber
\end{align}\end{linenomath}
Taking the averages of $G_{2,x,\text{jet}}$ and $G_{2,y,\text{jet}}$ over many events yields: 
\begin{linenomath}\begin{align}\label{eq:av}
&\langle G_{2,x,\text{jet}} \rangle = \langle Q_2\cos(2(\Psi_{\text{EP,}2}-\Psi_{\text{PP},2}))\rangle v_2^{\text{jet}} + \langle A_2 \rangle\\
&\langle G_{2,y,\text{jet}} \rangle = 0 \nonumber
\end{align}\end{linenomath}
In~(\ref{eq:av}) $v_2^{\text{jet}}$ has replaced the quantity $\langle \cos(2(\Psi_{\text{jet}}-\Psi_{\text{PP},2})) \rangle$.  It is assumed in this decomposition that $v_2^{\text{jet}}$ is independent of $Q_2$.  

Solving for $v_2^{\text{jet}}$ and $\Psi_{\text{EP,}2}$ requires calculating the higher moments of $\mathbf{G}_2$ listed in~(\ref{eq:G}).  Note that terms from higher mixed harmonics (on the order of $v_4^{\text{jet}}$ and above) are neglected in~(\ref{eq:G}):  
\begin{linenomath}\begin{align}\label{eq:G}
\langle G_{2,y,\text{jet}}^2 \rangle =& \tfrac{1}{2} \langle Q_2^2 \rangle \\
\langle G_{2,x,\text{jet}} G_{2,y,\text{jet}}^2 \rangle =& \tfrac{1}{4} \left(\langle Q_2^3\cos(2(\Psi_{\text{EP,}2}-\Psi_{\text{PP},2})) \rangle v_2^{\text{jet}} \right. \nonumber\\
&\left. \vphantom{v_2^{\text{jet}} }+ 2 \langle A_2 \rangle \langle Q_2^2 \rangle \right) \nonumber\\
\langle G_{2,y,\text{jet}}^4 \rangle =& \tfrac{3}{8} \langle Q_2^4 \rangle \nonumber
\end{align}\end{linenomath}

In~(\ref{eq:av}) the quantity $Q_2\cos(2(\Psi_{\text{EP,}2}-\Psi_{\text{PP},2}))$ is $Q_{2,x}$ in the participant plane frame.  It is necessary to assume functional forms for the distributions of $Q_{2,x,\text{PP}}$ and $Q_{2,y,\text{PP}}$ in order to solve the equations in~(\ref{eq:av}) and~(\ref{eq:G}).  The distributions of $Q_{2,x,\text{PP}}$ and $Q_{2,y,\text{PP}}$ are taken to be Gaussian with standard deviation $\sigma$.  The distribution of $Q_{2,x,\text{PP}}$ is centered at $\mu$ while the distribution of $Q_{2,y,\text{PP}}$ is centered at zero.  The relevant moments of $\mathbf{Q}_2$ are 
\begin{linenomath}\begin{align}\label{eq:moments}
\langle Q_{2,x,\text{PP}} \rangle &= \mu\\
\langle Q_2^2 \rangle &= \langle Q_{2,x,\text{PP}}^2 + Q_{2,y,\text{PP}}^2 \rangle = \mu^2 + 2\sigma^2 \nonumber\\
\langle Q_{2,x,\text{PP}}Q_2^2 \rangle &= \langle Q_{2,x,\text{PP}}^3 + Q_{2,x,\text{PP}}Q_{2,y,\text{PP}}^2 \rangle = \mu^3+4\mu\sigma^2 \nonumber\\
\langle Q_2^4 \rangle &= \langle (Q_{2,x,\text{PP}}^2 + Q_{2,y,\text{PP}}^2)^2 \rangle = \mu^4 + 8\mu^2\sigma^2 + 8\sigma^4\nonumber
\end{align}\end{linenomath}
The system of equations in~(\ref{eq:av})--(\ref{eq:moments}) can be solved to yield the parameters $\mu$ and $\sigma$:  
\begin{linenomath}\begin{align}\label{eq:sols}
&\mu^2 = \sqrt{8\langle G_{2,y,\text{jet}}^2\rangle^2-\tfrac{8}{3}\langle G_{2,y,\text{jet}}^4 \rangle}  \\
&\sigma^2 = \langle G_{2,y,\text{jet}}^2 \rangle - \tfrac{1}{2} \mu^2\nonumber
\end{align} \end{linenomath}
It is straightforward to solve for jet $v_2$ using Eqns.~(\ref{eq:av})--(\ref{eq:sols}).  

\subsection{Jet $v_2$}
The formula for $v_2^{\text{jet}}$ in this new method (denoted by $v_2^{\text{jet}}\{\text{QA}\}$) is given in Eqn.~(\ref{eq:v2jet}):  
\begin{linenomath}\begin{equation}\label{eq:v2jet}
v_2^{\text{jet}}\{\text{QA}\} = \frac{4\langle G_{2,x,\text{jet}} \rangle \langle G_{2,y,\text{jet}}^2 \rangle - 4\langle G_{2,x,\text{jet}} G_{2,y,\text{jet}}^2 \rangle}{\mu^3}
\end{equation}\end{linenomath}
Note that this method already accounts for the event plane resolution by calculating $\langle \cos(2(\Psi_{\text{jet}}-\Psi_{\text{PP},2})) \rangle$ instead of $\langle \cos(2(\Psi_{\text{jet}}-\Psi_{\text{EP,}2})) \rangle$.  It is not necessary to divide the resulting $v_2^{\text{jet}}\{\text{QA}\}$ by the event plane resolution.  

\subsection{Event Plane}
Additionally, Eqns.~(\ref{eq:av})--(\ref{eq:sols}) can be solved for $\langle A_2 \rangle$, the average shift in
$G_{2,x,\text{jet}}$ due to the jet.  
\begin{linenomath}\begin{equation}\label{eq:A}
\langle A_2 \rangle = \langle G_{2,x,\text{jet}} \rangle - \mu v_2^{\text{jet}}
\end{equation}\end{linenomath}
Once $\langle A_2 \rangle$ is known, it can be subtracted from $G_{2,x,\text{jet}}$ event by event.  Then the event plane can be calculated in a way that is, on average, not biased towards the jet, using Eqn.~(\ref{eq:EPnew}) which is based on Eqn.~(\ref{eq:EP}): 
\begin{linenomath}\begin{equation}\label{eq:EPnew}
\Psi_{\text{EP,2}} = \frac{1}{2} \tan^{-1} \left(\frac{\sum_i w_i\sin\left(2(\phi_i-\Psi_{\text{jet}})\right)}{\sum_i w_i\cos\left(2(\phi_i-\Psi_{\text{jet}})\right)-\langle A_2 \rangle}\right)+\Psi_{\text{jet}}
\end{equation}\end{linenomath}

\section{Higher Harmonics $v_n^{\text{jet}}$}
The quantities $G_{n,x,\text{jet}}$ and $G_{n,y,\text{jet}}$ can be evaluated at any order $n$: 
\begin{linenomath}\begin{align}
G_{n,x,\text{jet}} &= \sum_i w_i\cos(n(\phi_i-\Psi_{\text{jet}})) \\
&= Q_n\cos(n(\Psi_{\text{EP,}n}-\Psi_{\text{jet}})) + A_n\nonumber\\
G_{n,y,\text{jet}} &= \sum_i w_i\sin(n(\phi_i-\Psi_{\text{jet}})) \nonumber\\
&= Q_n\sin(n(\Psi_{\text{EP,}n}-\Psi_{\text{jet}}))\nonumber
\end{align}\end{linenomath}
However, beyond the definitions of $G_{n,x,\text{jet}}$ and $G_{n,y,\text{jet}}$, the method described above does not depend on the value of $n$.  Therefore the equations listed in~(\ref{eq:av})--(\ref{eq:v2jet}) can be used to solve for any $v_n^{\text{jet}}$.  Solving for the $n^{\text{th}}$-order event plane utilizes Eqns.~(\ref{eq:A}) and~(\ref{eq:EPnewn}), and is analogous to Eqn.~(\ref{eq:EPnew}): 
\begin{linenomath}\begin{equation}\label{eq:EPnewn}
\Psi_{\text{EP,}n} = \frac{1}{n} \tan^{-1} \left(\frac{\sum_i w_i\sin\left(n(\phi_i-\Psi_{\text{jet}})\right)}{\sum_i w_i\cos\left(n(\phi_i-\Psi_{\text{jet}})\right)-\langle A_n \rangle}\right)+\Psi_{\text{jet}}
\end{equation}\end{linenomath}

\section{Simulation}
This method has been tested on simple simulated events, each of which consists of a PYTHIA~\cite{pythia,pythia2} ($pp$-like) jet embedded in a thermal background.  PYTHIA jets are simulated in an interval of $\pm 2$ GeV around a given transverse energy $E_{\text{T}}^{\text{jet}}$.  The charged tracks in the PYTHIA event are analyzed with the anti-$k_\text{T}$ algorithm from the FastJet package~\cite{fastjet} (with a resolution parameter $R = 0.4$) to obtain the charged jet $p_{\text{T}}$ ($p_{\text{T}}^{\text{jet,ch}}$) and the angles of the jet axis ($\Psi_{\text{jet}}$,$\eta_{\text{jet}}$).  The charged jet $p_{\text{T}}$ from FastJet is required to satisfy $\frac{2}{3c}\left(E_{\text{T}}^{\text{jet}}-2\text{ GeV}\right) < p_{\text{T}}^{\text{jet,ch}} < \frac{2}{3c}\left(E_{\text{T}}^{\text{jet}}+2\text{ GeV}\right)$; this requirement ensures that at least one of the jets from PYTHIA is inside the detector acceptance.  Only tracks at midrapidity ($|\eta|<1$) are used in the simulation to mimic the acceptance of the STAR and ALICE TPCs~\cite{STARtpc,ALICEtpc}, and the jet axis is restricted to lie within $|\eta_{\text{jet}}| < 1-R$.  Note that the jet axis and $p_{\text{T}}$ are reconstructed prior to embedding the jet in the thermal background, to avoid issues related to full jet reconstruction in a heavy ion environment, although such a separation is not possible in experimental data.  The thermal ($T = 0.291$ GeV, which corresponds to approximately $\sqrt{s_{\text{NN}}} = 200$ GeV) background event is created such that the $2^{\text{nd}}$-order participant plane is correlated with $\Psi_{\text{jet}}$ according to Eqn.~(\ref{eq:fourierJet}) in order to produce a jet $v_2$.  The $p_{\text{T}}$- and centrality-dependent $v_2$ values of the background particles and the multiplicities of the background event are chosen to be consistent with observations from STAR~\cite{STARv2,STARmult} at RHIC.  

\subsection{Jet $v_2$}
\begin{figure*}[tb]
\centering
\includegraphics[width=0.9\linewidth, trim = 0mm 0mm 0mm 0mm, clip=true]{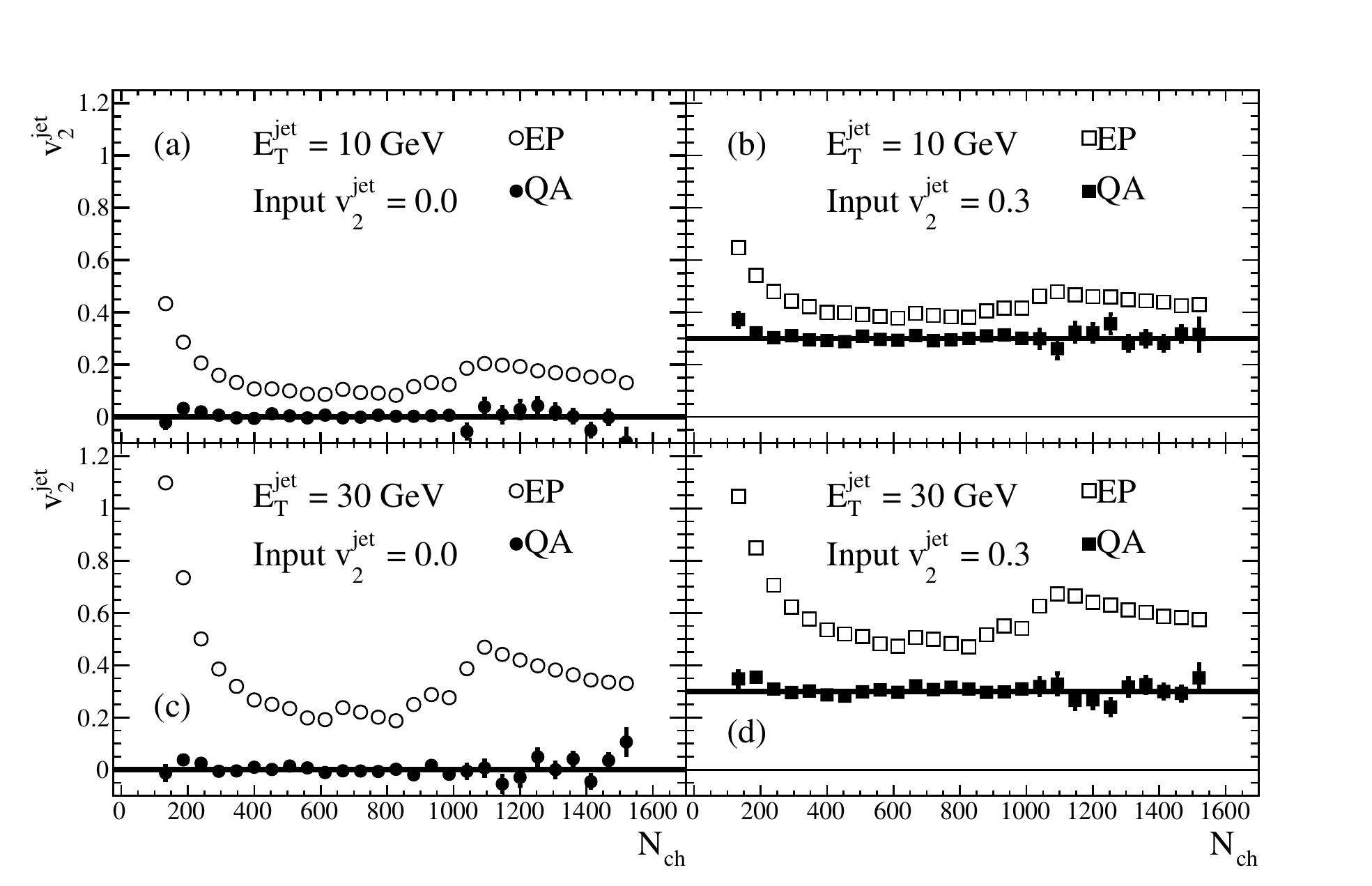}
\caption{\label{fig:jet10}Jet $v_2$ is calculated using the standard (EP) method (open symbols) and the new (QA) method (closed symbols) in simulated events consisting of a PYTHIA jet embedded in a thermal background ($T = 0.291$ GeV).  The simulated jet $v_2$ is denoted by a solid line for two cases: $v_2^{\text{jet}} = 0.0$ (left column) and $v_2^{\text{jet}} = 0.3$ (right column).  The results are shown for two jet energies: $E_{\text{T}}^{\text{jet}} = 10$ GeV (top row) and $E_{\text{T}}^{\text{jet}} = 30$ GeV (bottom row).  The $p_{\text{T}}$- and centrality-dependent $v_2$ values and multiplicities of the background are consistent with observations from STAR~\cite{STARv2,STARmult}.  Statistical errors are drawn but are often smaller than the symbol size.  The ``jumps'' in the distribution of $v_2^{\text{jet}}\{\text{EP}\}$ correspond to the edges of the centrality bins in~\cite{STARv2} that were used to simulate the bulk $v_2$.  }
\end{figure*}

Jet $v_2$ is calculated from all charged tracks in the combined event, without $p_{\text{T}}$-weighting.  The results are similar when $p_{\text{T}}$-weighting or a track $p_{\text{T}}$ cut is used.  For each bin in $N_{\text{ch}}$, the number of charged particles at mid-rapidity in the event, jet $v_2$ is calculated using the standard method (Eqns.~(\ref{eq:EP})--(\ref{eq:v2std})) and with the new method proposed here (Eqn.~(\ref{eq:v2jet})).  The results of the simulation are shown in Fig.~\ref{fig:jet10} for $v_2^{\text{jet}} = 0.0$ and $v_2^{\text{jet}} = 0.3$ in two cases: $E_{\text{T}}^{\text{jet}} = 10$~GeV and $E_{\text{T}}^{\text{jet}} = 30$~GeV.  

The results of the simulation indicate that the standard method of calculating $v_2^{\text{jet}}\{\text{EP}\}$ leads to an overestimation of the true jet $v_2$ by 5--20\% (or more, in peripheral collisions) due to the jet biasing the event plane calculation.  However, the new method of calculating $v_2^{\text{jet}}\{\text{QA}\}$ obtains accurate values of jet $v_2$.  The observed discrepancy between the simulated $v_2^{\text{jet}}$ and $v_2^{\text{jet}}\{\text{EP}\}$ is largest in the most central events where the bulk $v_2$ is low and the event plane is not well defined, and in the most peripheral events where the ratio of jet fragments to bulk particles is significant. The artificial jet -- event plane bias increases with jet energy.

\subsection{Event Plane}
\begin{figure*}[bt]
\centering
\includegraphics[width=0.9\linewidth, trim = 0mm 0mm 0mm 0mm, clip=true]{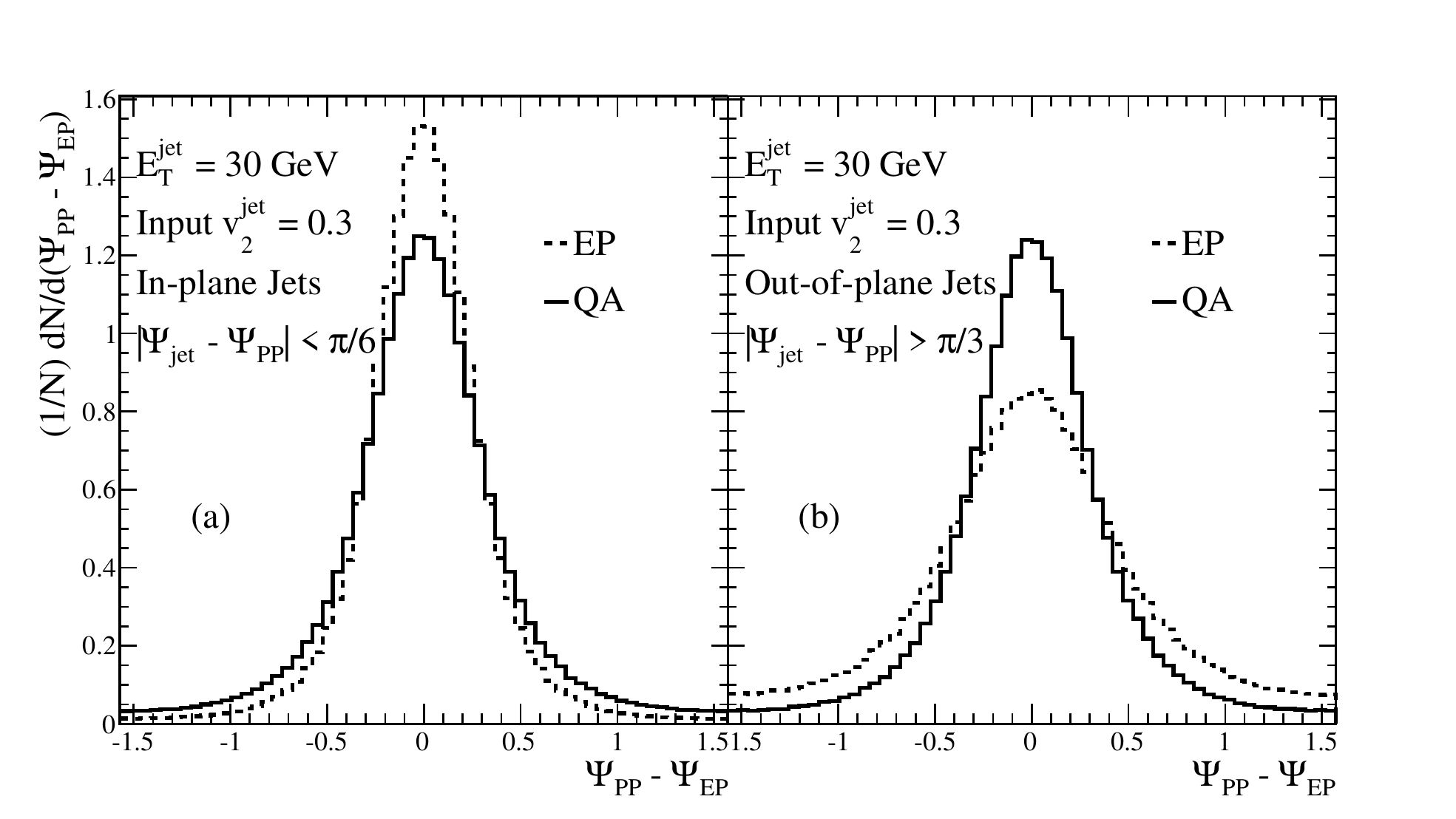}
\caption{\label{fig:EP} The event plane is calculated using the standard (EP) and new (QA) methods in simulated events consisting of a PYTHIA jet with $E_{\text{T}}^{\text{jet}} = 30$ GeV embedded in a thermal background ($T = 0.291$ GeV).  The $p_{\text{T}}$- and centrality-dependent $v_2$ values and multiplicities of the background are consistent with observations from STAR~\cite{STARv2,STARmult}, and $v_2^{\text{jet}} = 0.3$.  The difference between the reconstructed event planes and the participant plane are folded into the interval $\left(-\frac{\pi}{2},\frac{\pi}{2}\right)$ and are shown for two cases: (a) the jet is aligned with the participant plane and (b) the jet is perpendicular to the participant plane.  The results are integrated over $N_{\text{ch}}$.}
\end{figure*}

The event plane was calculated using the standard method (Eqn.~(\ref{eq:EP})) and the new method (Eqn.~(\ref{eq:EPnew})) and then compared with the simulated participant plane in Fig.~\ref{fig:EP}.  Both methods accurately reconstruct the participant plane, and the results are shown separately for the cases in which the jet is in-plane ($|\Psi_{\text{jet}} - \Psi_{\text{PP},2}| < \pi/6$) and out-of-plane ($|\Psi_{\text{jet}} - \Psi_{\text{PP},2}| > \pi/3$).  It is clear from Fig.~\ref{fig:EP} that the standard event plane resolution depends on the orientation of the jet with respect to the participant plane.  When the jet is aligned with the participant plane, the resolution of the standard event plane method is high because the jet pulls the event plane towards the participant plane.  When the jet is perpendicular to the participant plane, the standard event plane is pulled away from the true participant plane, thus lowering the resolution.  However, the resolution of the event plane calculated with the new (QA) method does not depend on the orientation of the jet to the participant plane, as is desirable.  

Table~\ref{tab:sigmas} shows the standard deviation ($\sigma$) of the event plane from the participant plane, for both the standard method and the new (QA) method, in two centrality bins and for three orientations of the jet to the participant plane.  The results show that while the standard event plane resolution depends on the orientation of the jet, the QA event plane resolution is constant.  Furthermore, the resolution of the QA event plane does not differ significantly from the resolution of the standard event plane in either of the centrality classes.  

\begin{table*}[ht]
\centering
\begin{tabular}{|ccc|}
\hline
Jet -- Participant Plane Orientation & $\sigma_{\text{EP}}$ & $\sigma_{\text{QA}}$ \\ \hline
$0 < |\Psi_{\text{jet}} - \Psi_{\text{PP},2}| < \pi/6$ &  $0.3419\pm0.0004$ & $0.4378\pm0.0004$ \\
$\pi/6 < |\Psi_{\text{jet}} - \Psi_{\text{PP},2}| < \pi/3$ &  $0.4461\pm0.0004$ & $0.4374\pm0.0004$ \\
$\pi/3 < |\Psi_{\text{jet}} - \Psi_{\text{PP},2}| < \pi/2$ &  $0.5755\pm0.0006$ & $0.4397\pm0.0006$ \\ \hline\hline
Centrality & $\sigma_{\text{EP}}$ & $\sigma_{\text{QA}}$ \\ \hline
Peripheral ($107 \leq N_{\text{ch}} < 800$) & $0.3426\pm0.0004$ & $0.3399\pm0.0004$ \\
Central ($800 \leq N_{\text{ch}} \leq 1546$) & $0.4912\pm0.0004$ & $0.5157\pm0.0003$ \\ \hline
\end{tabular}
\caption{\label{tab:sigmas} Standard deviation of the event plane from the participant plane, when the event plane is calculated with the standard method ($\sigma_{\text{EP}}$) and with the new method ($\sigma_{\text{QA}}$).  The results are shown for two centrality bins, and three orientations of the jet to the participant plane.  Errors are statistical only.}
\end{table*}

\subsection{Jet $v_3$}
\begin{figure*}[bt]
\centering
\includegraphics[width=0.9\linewidth, trim = 0mm 63mm 0mm 0mm, clip=true]{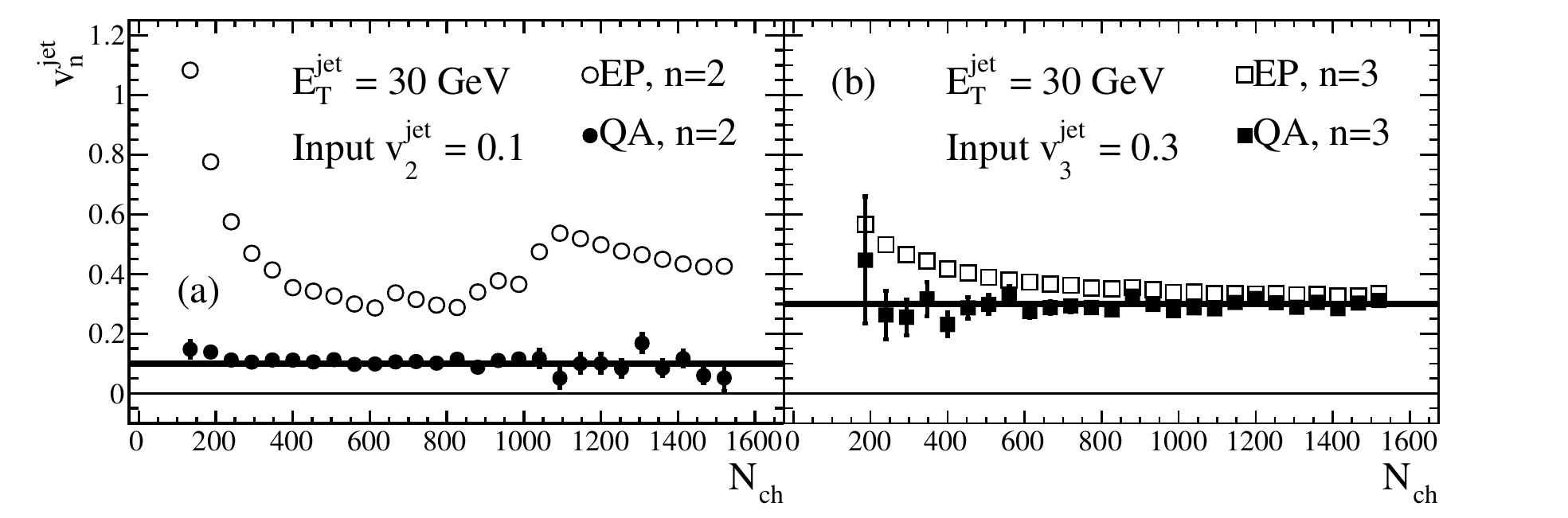}
\caption{\label{fig:jetv3}Jet $v_2$ and $v_3$ are calculated simultaneously using the standard (EP) method (open symbols) and the new (QA) method (closed symbols) in simulated events consisting of a PYTHIA jet with $E_{\text{T}}^{\text{jet}} = 30$ GeV embedded in a thermal background ($T = 0.291$ GeV).  The $p_{\text{T}}$- and centrality-dependent $v_2$ values and multiplicities of the background are consistent with observations from STAR~\cite{STARv2,STARmult}.  Solid lines denote the simulated (a) jet $v_2 = 0.1$ and (b) jet $v_3 = 0.3$.  Statistical errors are drawn.}
\end{figure*}

The simulation was modified to include a background modulated by $v_2$ and $v_3$ and a jet correlated to $\Psi_{\text{PP,}2}$ and $\Psi_{\text{PP,}3}$ via $v_2^{\text{jet}}$ and $v_3^{\text{jet}}$.  The underlying event $v_3$ was chosen to equal the $v_2$ measured by STAR in the 5--10\% centrality bin~\cite{STARv2}.  The $2^{\text{nd}}$- and $3^{\text{rd}}$-order participant planes are each correlated with the jet axis, but are not explicitly correlated to each other.  Figure~\ref{fig:jetv3} shows the results of a simulation in which 30 GeV jets are chosen to have $v_2^{\text{jet}} = 0.1$ and $v_3^{\text{jet}} = 0.3$.  The new jet $v_n$ method can accurately measure the jet $v_2$ and $v_3$ harmonics independently.  The standard method estimation of jet $v_3$ is less biased than the estimation of $v_2^{\text{jet}}$, because in the case of odd harmonics, the two jets in a dijet pair pull the event plane in opposite directions, and thus the bias is largely canceled.  

\subsection{Jets at the LHC}
\begin{figure*}[bt]
\centering
\includegraphics[width=0.45\linewidth, trim = 0mm 63mm 10cm 0mm, clip=true]{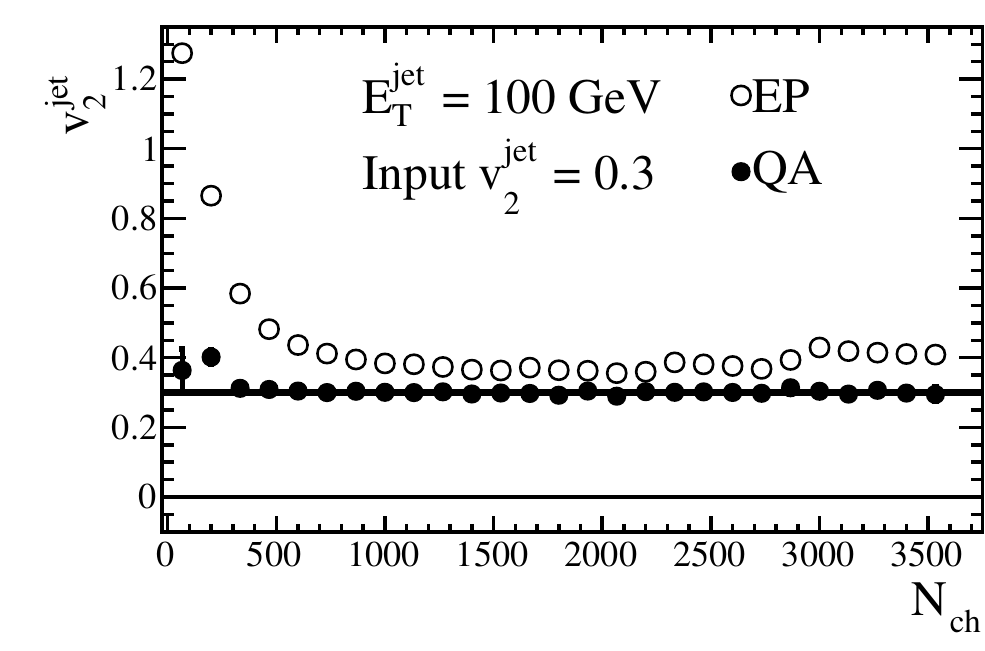}
\caption{\label{fig:ALICE}Jet $v_2$ is calculated using the standard (EP) method (open symbols) and the new (QA) method (closed symbols) in simulated events consisting of a PYTHIA jet with $E_{\text{T}}^{\text{jet}} = 100$ GeV embedded in a thermal background ($T = 0.350$ GeV).  The $p_{\text{T}}$- and centrality-dependent $v_2$ values and multiplicities of the background are consistent with observations from ALICE~\cite{ALICEv2,ALICEmult}.  The simulated jet $v_2$ is 0.3 and is denoted by a solid line.  Statistical errors are smaller than the symbol size.}
\end{figure*}

The simulation was also modified so that the $v_2$ values of the background particles and the multiplicities of the thermal ($T = 0.350$ GeV) background event are consistent with observations from ALICE~\cite{ALICEv2,ALICEmult} at the LHC.  The results are shown in Fig.~\ref{fig:ALICE} for $E_{\text{T}}^{\text{jet}} = 100$ GeV and $v_2^{\text{jet}} = 0.3$.  Due to the higher multiplicities at LHC energies, the standard event plane method does not overestimate jet $v_2$ as drastically as at RHIC energies.  The new method also successfully recovers the correct jet $v_2$ at LHC energies.  

\section{Discussion \& Conclusions}
As shown by the above simulation study, this new method can accurately calculate jet $v_2$ and the $2^{\text{nd}}$-order event plane in a simple picture of a jet embedded in a heavy ion background exhibiting elliptic flow.  The method also works for higher-order anisotropies.  

This method is tested in a simulation which does not include medium-induced jet modification.  The decomposition in Eqn.~(\ref{eq:G0}) implies that the bulk particle distribution is independent of the jet.  However, if the jet is modified by interactions with the medium, then the bulk must be modified to some degree as well.  It is expected that jet-medium interactions would produce structures in the azimuthal distribution of particles that are correlated with (or symmetric about) the jet axis, and therefore jet quenching would be reflected in the event-by-event magnitude of $\mathbf{A}_n$, rather than in $\mathbf{Q}_n$, to first order.  

This method does not assume anything about the magnitude of the jet vector ($A_n$) or about the $v_n$ components of the underlying event.  However, the equations for calculating $v_n^{\text{jet}}$ and $\Psi_{\text{EP},n}$ rely on the assumption that the fluctuations of $Q_{n,x}$ and $Q_{n,y}$ in the participant plane frame can be described by Gaussian distributions.  This assumption follows from Ref.~\cite{gaussianQ}, although it is noted that there are small deviations from the Gaussian ansatz in peripheral collisions.  However, Ref.~\cite{flowfluc} disputes this assumption, stating that a Gaussian distribution is a poor description of event-by-event fluctuations in a Monte Carlo Glauber model for all but the most central collisions.  This new jet $v_n$ method may be generalized to account for different distributions of $Q_{n,x,\text{PP}}$ and $Q_{n,y,\text{PP}}$ by rederiving Eqns.~(\ref{eq:moments})--(\ref{eq:EPnew}) with alternative non-Gaussian functional forms.  

Although the method presented here assumes perfect resolution of the jet axis, accounting for the jet axis resolution in the measurement of $v_2^{\text{jet}}$ does not require significant changes to the equations.  This method is improved when it is applied in fine bins in centrality (multiplicity) and jet energy, in order for the average quantities ($\langle A_n \rangle$, $\langle Q_n \rangle$, etc) to be meaningful.  Unfortunately, high statistics are required in order to obtain reasonable results.  For this reason the application of this method may not be feasible at RHIC, whereas statistics may be sufficient at the LHC.  

While jets can influence event plane and $v_2$ measurements, the production of such high-$p_{\text{T}}$ jets is rare and therefore only a small subset of events are affected in most event plane analyses.  Furthermore, in $v_2$ analyses of objects that are uncorrelated to jets, the jet -- event plane bias only contributes to a small decrease in the event plane resolution.  It is only when calculating the $v_2$ of jets themselves, or particles likely to come from jet production (such as high-$p_{\text{T}}$ hadrons), that the jet can cause a significant overestimation in the $v_2$ calculation.  

A measurement of jet $v_n$ for odd $n$ could potentially yield more information about jet quenching than the even $v_n^{\text{jet}}$ harmonics.  Since parton pairs produced in hard scatterings are essentially symmetric under an azimuthal rotation by $\pi$, and odd harmonics are antisymmetric under the same rotation, it is not possible for jet production to have any intrinsic correlation with an odd event plane.  Barring detector acceptance effects, jet reconstruction algorithms will likely find (or assign a higher energy to) the jet in a dijet pair which undergoes less modification.  If a nonzero odd $v_n^{\text{jet}}$ term is measured, it could be indicative of the correlation between the $n^{\text{th}}$-order event plane and jets which undergo less modification than their recoiling partners, illustrating pathlength-dependent medium-induced jet modification.  

In conclusion, the calculation of jet $v_n$ can be beneficial for studies of medium-induced jet modification.  However, reconstructing the event plane in the presence of a jet is non-trivial due to the effects of jet fragments on the event plane calculation.  The artificial jet -- event plane bias that arises when jet particles are included in the event plane calculation can lead to a significant overestimation of jet $v_2$.  The method presented here utilizes knowledge of the jet axis from full jet reconstruction to accurately calculate jet $v_2$ and $v_3$ as well as an event plane that is, on average, unbiased by the presence of a jet.  

\begin{acknowledgments}
The author would like to thank Art Poskanzer, J\"{o}rn Putschke, Paul Sorensen, Sergei Voloshin, and the members of the Yale Relativistic Heavy Ion Group for their helpful discussions.  

This work is supported in part by the facilities and staff of the Yale University Faculty of Arts and Sciences High Performance Computing Center. This research is supported in part by the Department of Energy Office of Science Graduate Fellowship Program (DOE SCGF), made possible in part by  the American Recovery and Reinvestment Act of 2009, administered by ORISE-ORAU under contract no. DE-AC05-06OR23100.
\end{acknowledgments}

\bibliography{biblio}{}
\bibliographystyle{apsrev4-1}
\end{document}